\documentclass[sn-mathphys]{sn-jnl}

\jyear{2022}%

\theoremstyle{thmstyleone}%
\theoremstyle{thmstyletwo}%

\theoremstyle{thmstylethree}%

\usepackage{makecell}
\begin{document}

\title[``Explanation'' is Not a Technical Term]{``Explanation'' is Not a Technical Term: The Problem of Ambiguity in XAI}


\author*[1]{\fnm{Leilani H.} \sur{Gilpin}}\email{lgilpin@ucsc.edu}
\author[2]{\fnm{Andrew R.} \sur{Paley}}\email{andrewpaley@northwestern.edu}
\author[2]{\fnm{Mohammed A.} \sur{Alam}}\email{mohammed.alam@northwestern.edu}
\author[2]{\fnm{Sarah} \sur{Spurlock}}\email{sarah.spurlock@northwestern.edu}
\author[2]{\fnm{Kristian J.} \sur{Hammond}}\email{kristian.hammond@northwestern.edu}

\affil*[1]{\orgdiv{Department of Computer Science \& Engineering}, \orgname{University of California, Santa Cruz}, \orgaddress{\street{1156 High Street}, \city{Santa Cruz}, \postcode{95060}, \state{CA}, \country{USA}}}

\affil[2]{\orgdiv{Department of Computer Science}, \orgname{Northwestern University}, \orgaddress{\street{633 Clark Street}, \city{Evanston}, \postcode{60208}, \state{IL}, \country{USA}}}



\abstract{There is broad agreement that Artificial Intelligence (AI) systems, particularly those using Machine Learning (ML), should be able to ``explain'' their behavior. Unfortunately, there is little agreement as to what constitutes an ``explanation.''  This has caused a disconnect between the explanations that systems produce in service of explainable Artificial Intelligence (XAI) and those explanations that users and other audiences actually need, which should be defined by the full spectrum of functional roles, audiences, and capabilities for explanation. In this paper, we explore the features of explanations and how to use those features in evaluating their utility. We focus on the requirements for explanations defined by their functional role, the knowledge states of users who are trying to understand them, and the availability of the information needed to generate them. Further, we discuss the risk of XAI enabling trust in systems without establishing their trustworthiness and define a critical next step for the field of XAI to establish metrics to guide and ground the utility of system-generated explanations.}




\maketitle

\section{Introduction}\label{introduction}
Artificial Intelligence (AI) is everywhere. Embodied in personal assistants, medical diagnosis, facial recognition, game playing, and autonomous driving, these technologies already touch a huge and diverse audience; an audience that, whether a system is right or wrong, wants to see explanations as well as answers\footnote{A suggested gold standard is to only use interpretable models~\cite{rudin2018please}.   We agree that, all things being equal, it is favorable to use an interpretable model rather than an opaque one. Depending on the task, however, it is sometimes not possible or even desirable to utilize an interpretable model.  For example, in complex systems like self-driving cars, the ``opaqueness'' could be due to the use of a proprietary model or the need to support real time responsiveness.  Replacing such a mechanism with an interpretable model is not feasible for the domain or the functionality. }.

Providing these explanations is difficult. Most modern AI models are constructed using systems that are opaque, making the details of their performance virtually unknowable. This problem is amplified in that even when an explanation for a system can be generated, it may be ill-suited to the needs of the user of the system. It may very well be an ``explanation,'' just not one that is useful.

While there are certainly technical problems involved with generating explanations of Machine Learning (ML) models, they are overshadowed by the problem of the lack of agreement as to what we mean by the term. This lack of agreement, this ambiguity, has resulted in a situation in which ``explanations'' are being built that may fit a technical definition of the term but do not serve the purpose for which they are designed.

Research in explanation is focused on mechanisms developed to compensate for the opaque or “black box” nature of algorithms like deep neural networks (DNN). The DARPA program on explainability~\cite{gunning-xai}, is intended to develop explanations or model-dependent reasons leading up to a DNN prediction. For example, ``the DNN classified this image as a ‘cat’ because it has `cat-like' ears and whiskers.'' This has led to a new area of research on model distillation and extrinsic explanation approaches like SHAP~\cite{shap}, LIME ~\cite{why-trust}, and saliency maps~\cite{grad-cam}, which aim to approximate complicated black box predictions into simpler approximations. While these methods can be useful for model debugging~\cite{sanity-checks}, they can be ``fooled'' and perturbed~\cite{slack2020fooling} just like the underlying opaque model~\cite{fooled}. 

Recently, it has been shown that explanations are a ``false hope'' for building trust in ML for healthcare ~\cite{ghassemi2021false}.  These XAI methods results are difficult to understand; as they are approximations for complex, opaque decision-making systems.  Further, explanations are not sufficient, especially when ``explaining'' why the underlying model prediction is wrong in the healthcare domain~\cite{jacobs2021machine}.  The problem is that explainability is not a well-defined goal: there is no common definition, metrics, or benchmarks for success.  An utterance may meet some technical definition of ``explanation'' but be different from the role it is supposed to play and not aligned to what users may expect or require.  The term ``explanation'' is a classic example of what Marvin Minsky referred to as suitcase words: words that contain multiple meanings which are interpreted in different ways for different people in different contexts~\cite{emotion-machine}.

Researchers, users, and domain experts, etc. often move between technical and colloquial senses of the word interchangeably. AI researchers promote their work as an explanation, also shifting back and forth between the technical and colloquial. The result is the growth of a set of approaches that provide what are, in fact, explanations, just not the ones we need.


\begin{figure}
    \centering
    \includegraphics[width=\textwidth]{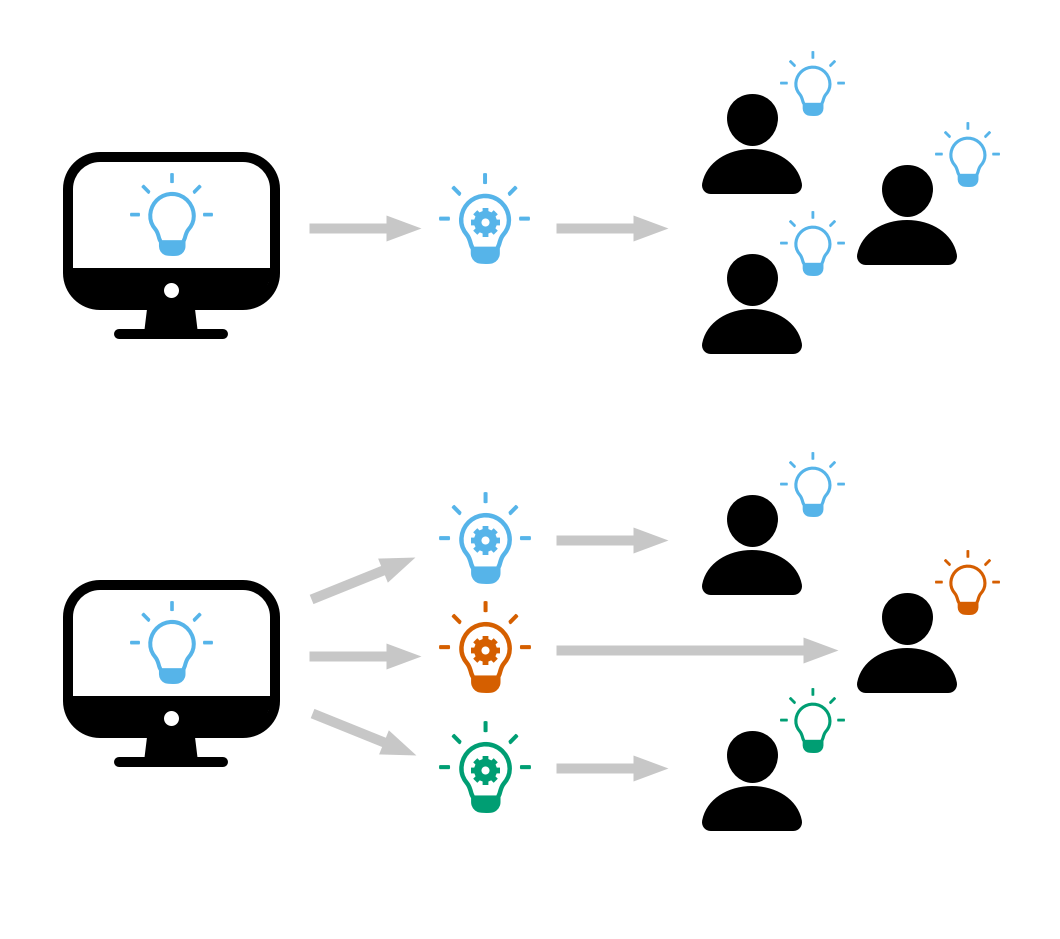}
    \caption{A monolithic versus functional view of explanation.  The gear represents the interpretation of an explanation which is displayed to a user.}
    \label{fig:fig1}
\end{figure}

Rather than taking a view of explanations as undifferentiated artifacts shared by multiple users, we view them as generated in response to their functional roles, audience, and data access. Analysis of these features allows us to determine if the different explanations satisfy their functional goals:
\begin{enumerate}
    \item \textbf{Functional role}: How is the explanation going to be used?
    \item \textbf{Audience}: To whom is it directed and what is their knowledge of the system and domain?
    \item \textbf{Capabilities}: What are the capabilities of the system constructing the explanation and the source of data/knowledge used to do so?
\end{enumerate}
In this paper, we will unpack these issues and show how a one-size-fits-all explanation approach is not sufficient (see Figure~\ref{fig:fig1}). We will review the explanations generated by current XAI algorithms and ask if they are aligned with their stated functional roles. We will also discuss the difference between \emph{trusted} and \emph{trustworthy} and how the distinction impacts XAI.

\section{Motivation}
Explanations are often intended to satisfy a wide range of functional roles. For developers, they can give guidance in the debugging process.  For users, they can provide justifications for a system’s decisions, or insight into features that might be adjusted to change outcomes. And for regulators and other legal entities, they can provide audit trails that can be used to determine responsibility.

Consider a self-driving car that runs itself into a wall to avoid a pedestrian:
\begin{itemize}
    \item A developer may want to know how the inputs were interpreted to refine the system. That is, an explanation that describes ``how'' the decision was made.
    \item  The car’s passenger or owner may need to know that it was trying to save human life, the functional ``why.''
    \item  The courts may want to know who is responsible for the property damage and get an explanation at the level of the rules that were used to make this decision in order to track down the ``cause'' underlying the decision.
\end{itemize}

\noindent
Each stakeholder requires an explanation that fits the functional role but explanations that serve other roles are useless to them. Developers, user, and courts each have different information needs that are served by different types of explanation. 

\vspace{1em}
\noindent\fbox{%
    \parbox{\textwidth}{%
        An explanation’s functional role defines the information that needs to be communicated.
    }%
}
\vspace{1em}

\noindent
Because explanations are links between agents, we also need to consider the audience and their knowledge background. An explanation of the rationale for a diagnosis based on X-ray data that is provided to a radiologist (focused on visual features) is different than one provided to other physicians (focused on interpretation), which is also different than one provided to a developer (specific weights and activation flows). Different users have different information needs and different knowledge backgrounds. This means that different users need different types of explanations.


\vspace{1em}
\noindent\fbox{%
    \parbox{\textwidth}{%
        The state of the audience’s initial knowledge of both the domain and processes define a second set of requirements related to the detail and vocabulary used in the explanation.
    }%
}
\vspace{1em}

\noindent
On the other side of the communication flow is the machine itself. The knowledge state of the machine and how that knowledge is used to generate explanations define the capabilities of the system. A system answering questions based on aggregated evidence (e.g., IBM Watson) has access to the rules that support a particular answer. A system that uses a DNN to recognize faces, has access to the equations linked to the individual nodes. And a system that utilizes decision trees or Naïve Bayes, has access to the rules at each node and the probabilities linking input features to predicted outcomes. 

\vspace{1em}
\noindent\fbox{%
    \parbox{\textwidth}{%
A system’s capabilities and access to knowledge about its own reasoning determine the scope and validity of the explanations it can generate.
    }%
}
\vspace{1em}

\noindent
These three factors define the basis for metrics for an evaluation calculus that can be used to evaluate an explanation based on whether it serves the right functional role with the right level of elaboration for its audience supported by the system’s knowledge of its own reasoning.

By unpacking the idea of ``explanation'' into these factors, XAI can go beyond the ``checking the box'' phase to one in which explanations can play the role for which they were designed. Unless we can evaluate these artifacts from the perspective of purpose, we will remain trapped in the world of explanations in name only. 



To understand, develop, and evaluate machine generated explanations, we need to unpack these central defining features of the explanations, the systems that build them and the users who need to understand them.

\subsection{The Requirements: Functional Roles}
Explanations are communication tools that are used in a variety of functional roles. They serve to help debug systems, support human decision making, justify machine decisions, and even give comfort to users. In efforts to define explanation precisely, recent research in XAI has looked to the social sciences for guidance~\cite{miller2019explanation}. Unfortunately, one of the early findings in the realm of explanation was the discovery that the human bar on what constitutes a ``good explanation'' is shockingly low. As Langer discovered in 1978, people are satisfied with ``Placebic'' information if it has the syntactic form of an explanation~\cite{langer1978mindlessness}. This work demonstrated that we need an approach to evaluating explanations that goes beyond human assessment and acceptance.

To illustrate the importance of functional roles and users, take the field of medicine: a domain rife with nascent AI applications and an array of potential users with varying skill sets, domain knowledge and technical capabilities. This eco-system includes the data scientists and engineers who work to develop capabilities and train models for downstream tasks, the medical practitioners who interface with these systems and the patients themselves who are the true stakeholders. 


For the \textbf{engineers and developers}, explanations are close to the machinery, and the shared language is technical and can contain machine representation. These users have experience working with models, understand limitations, can run experiments, and can intuit from incomplete or partial ``explanations'' as they perform their task: debugging and iterating on the model to improve its performance and make it trustworthy.

For the \textbf{doctors}, explanations are less about the core mechanics of the models themselves and reflect the logic of the domain, affording exploration, counterfactuals, cohort comparison and transparent reasoning about the features most pertinent to a diagnosis. For example, if a system provides a warning that a particular patient is showing signs of possible heart failure, a doctor will want a list of relevant factors in order of importance or concern, as well as the patient’s prognosis compared to that of other patients, and how that prognosis changes if certain factors are amended. At their best, the interaction between a doctor and an intelligent system should seek to mirror the sorts of interactions two doctors might have when collaborating on the task of developing a diagnosis and refining a treatment plan. 

\textbf{Patients} have very different requirements. In general, a patient’s goal is more immediate and comes with higher personal stakes. This is the realm of personal decision making and, in cases of emergency, immediate action. These individuals are less informed about aspects of the medical domain, and systems tailored to them must account for that information asymmetry. Thus, XAI, in this use case, becomes less about collaboration or justification and instead is geared towards confidence building, risk assessment, contextualization, and guided calls to action 

Finally, \textbf{regulators and auditors} require explanations that scope across the mechanisms of a system, the data used to train it, and the medical practices that it embodies. More so, they require explanations that might include elements of the human in the loop in order to determine responsibility and culpability.

These scenarios uncover some basic functional roles. These functionalities are related to the development and real-time utilization of a system. There are others that are more closely tied to the retrospective evaluation of past performance and aggregate behavior. The point here is simple: When we design explanations, we must consider the role that they play in the overall use of the systems. 

\begin{itemize}
    \item For engineers building a system, explanations need to touch on the aspects of a system’s decisions that can be used in debugging, referencing the data, feature selection, and comparisons.
    \item For users interpreting the recommendations of a system, explanations need to include features that can be used to support exploration of hypotheticals, counterfactuals, cohort comparison and likelihoods.
    \item For stakeholders impacted by a decision, explanations need to support trust and confidence building, risk assessment, contextualization, and decision support.
    \item For auditors and regulators, explanations need to support comparisons and aggregate review of performance and the trail of both algorithmic and human decisions that led to it. 
Each of these different functional roles defines a set of requirements that we can use to evaluate explanations
\end{itemize}
Each of these different functional roles defines a set of requirements that we can use to evaluate explanations.

\begin{figure}
    \centering
    \includegraphics[width=\textwidth]{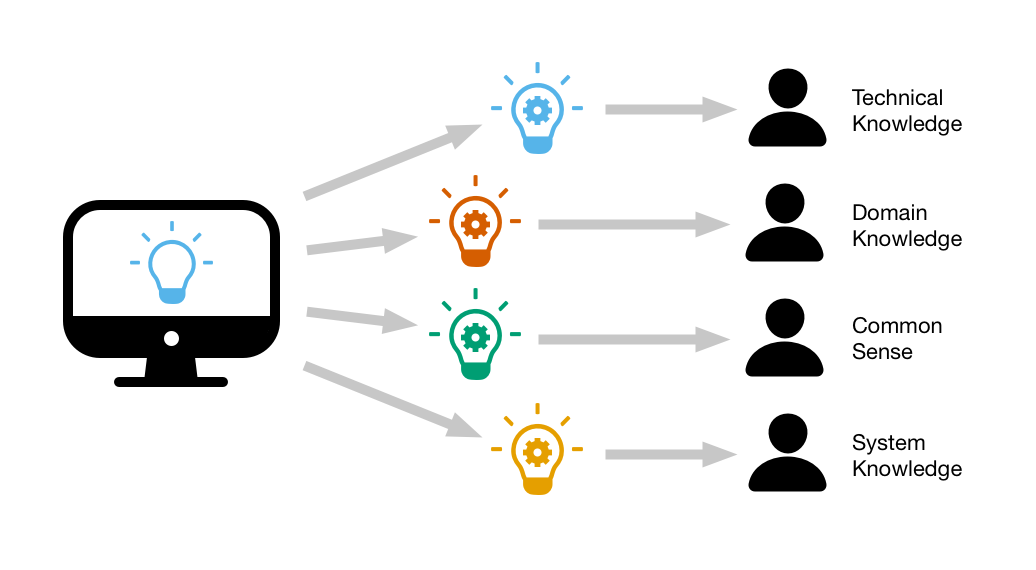}
    \caption{The types of user knowledge required to interpret machine explanations for engineers/scientists, domain experts, and impacted stakeholders.}
    \label{fig:user-roles}
\end{figure}

\subsection{The Requirements: User Knowledge}

The functional roles that explanations play in the development, use, and evaluation of systems define one set of requirements. The knowledge state of the various stakeholders interacting with the system defines another. This knowledge can be characterized in a variety of ways, but we will focus here on knowledge related to the technologies of ML and data and the domains of application areas. 


Model engineers and data scientists are at one end of the spectrum – their assessment of a system is based on an informed judgment of a model’s capabilities and correctness at both technical and domain levels. As noted above, the functional role of XAI starts in the realm of model debugging and diagnostics – a tool set aimed at technical users capable of deciphering machine representations, interpreting model performance metrics, and updating model training to improve performance. The assumption is that these engineers and data scientists have working knowledge of the technologies and can understand explanations as this level.

Domain experts may not need explanations that provide insight into the technical workings of systems. They want and understand explanations at the domain level. While it is often the case that they have the right level of domain knowledge, it is always a possibility that they lack detailed knowledge of specific domain level features and ideas.

Impacted stakeholders are highly variable: they may have no knowledge of either the domain or the technology. They do not have expertise in either but do have basic knowledge of how the world works. 

Stakeholders such as auditors and regulators have specialized knowledge of the ways in which data and algorithms interact and how to look at the performance of a system through the lens of comparison and systemic issues.  This is visualized in Figure~\ref{fig:user-roles}. 


\section{System Capabilities}
In philosophy, an explanation’s functional role is to answer a ``why question''~\cite{sylvain}. In Human-Computer Interaction (HCI), an explanation’s functional role is to inform the user. In the AI and ethics subfield an explanation’s functional role is to be an ``inquirer,'' where ``intelligibility'' is noted as a user-defined goal~\cite{zhou2020different}. For most developers, however, explanations are technical debugging tools that don’t transfer well into different uses. 

Regardless of the requirements of the target purpose and user of an explanation, the question on the other side of the equation is whether systems have the capabilities and knowledge needs to satisfy them. 

As with the examination of requirements, the notion of system capability can be unpacked into specific areas: interpretability, accuracy, elaborateness, and faithfulness.  These atomic capabilities are a starting point for understanding and interpreting XAI performance.

Each of these capabilities can be viewed through the lens of the target requirements defined by the functional roles and the knowledge levels of the various stakeholders.

\begin{itemize}
    \item \textbf{Interpretability} is how understandable the output representation is to the audience. This depends on the target audience and the task. For example, a saliency map that highlights part of a radiology image that was ``important'' to classifying it as cancerous is understandable to a doctor, but may not be as understandable to a patient without subject matter expertise. A saliency map is an approximate representation of a DNN that may be useful for model debugging by system experts, if the map is an accurate representation~\cite{different-intelligibility}. However, Adebayo et al showed that these visualizations may be misleading~\cite{sanity-checks}.
    \item \textbf{Accuracy} for an explanation is based on correctness, except that it is in terms of the explanation itself. This is dependent on both the domain and the user. For example, saliency maps, and other approximate methods, do not have high accuracy. A saliency map applied to cancer images may highlight the hospital name, indicating that the reasons supporting the diagnosis is the hospital where the image was taken~\cite{leakage}. This explanation is true to the model but the model is faulty. The explanation is accurate but the model is not.
    \item \textbf{Elaborateness} is the depth or detail of the explanation. For example, a doctor may tell a patient that their diagnosis looks ``similar to'' another diagnosis. This is a ``shallow'' explanation; it does not point to the root cause of the diagnosis. An explanation that points to the root cause is elaborate. For example, if the doctor tells a patient why the diagnosis looks similar, then the elaborative explanation is useful for the patient’s benefit. The need for elaboration depends on functional role and user.
    \item \textbf{Faithfulness} is how well the explanation describes the underlying model. This is also known as model completeness~\cite{explaining-explanations}. For example, a saliency map that highlights the part of a radiology image that was ``important'' to classifying it as cancerous is an approximate method, so it is not complete or faithful to the underlying model’s processing. 
\end{itemize}

While the mechanisms underlying opaque machine learning (ML) models may be well understood, how they pick up on cues from training data is seldom transparent. Simple ML models may not pose much of a problem with transparency. But as parameter or algorithmic complexity grows, the model becomes more challenging to understand. The result is an opaque system that neither explains itself nor divulges information in the form of human-interpretable explanations, falling short of the XAI capabilities that could potentially tease out model shortcomings. 

\subsection{Trusted but not Trustworthy: A New Dark Pattern}
One of the dominant themes in XAI is the notion of trust. We need to be able to trust the systems that we use. A problem with this characterization of the goal is that it is focused on the user rather than the system.  The focus is on getting users to trust a system rather than on making the system trustworthy. 

Such trust is developed through output assessment over time and through a host of factors external to the model output itself (including the apparent trust of other experts, design decisions at the system and user experience levels, and ancillary components such as domain-informed conversational interfaces). Further, this sort of trust is hard won and easily lost. Thus, the broad goal of trusted XAI is potentially far more complex and is likely to be less generalizable and more context-dependent, as research into stakeholders' needs in the space of XAI has shown~\cite{bhatt2020machine}.

As with people, it’s possible for a system to create a veneer of trust – to become trusted without being trustworthy. In XAI, we can think of this as a new form of “dark pattern,” a term borrowed from user experience design for design patterns that are aimed at tricking a user into behaviors that don’t necessarily align with their goals. While it’s reasonable to hope most practitioners in the space of XAI will have better intentions, the potential negative impact of intelligent systems that inspire trust but do so through incomplete, inaccurate or nefarious means is significant.

This issue scopes beyond the individual to organizations. Which is to say, it is possible to build systems that are exceptionally well trusted because of what amounts to marketing strategies rather than their fundamental trustworthiness. In much the same way that individual assessments of systems regarding trust can be manipulated, organizational assessments can be manipulated. A more rigorous approach must be taken.

\begin{figure}
    \centering
    \includegraphics[width=\textwidth]{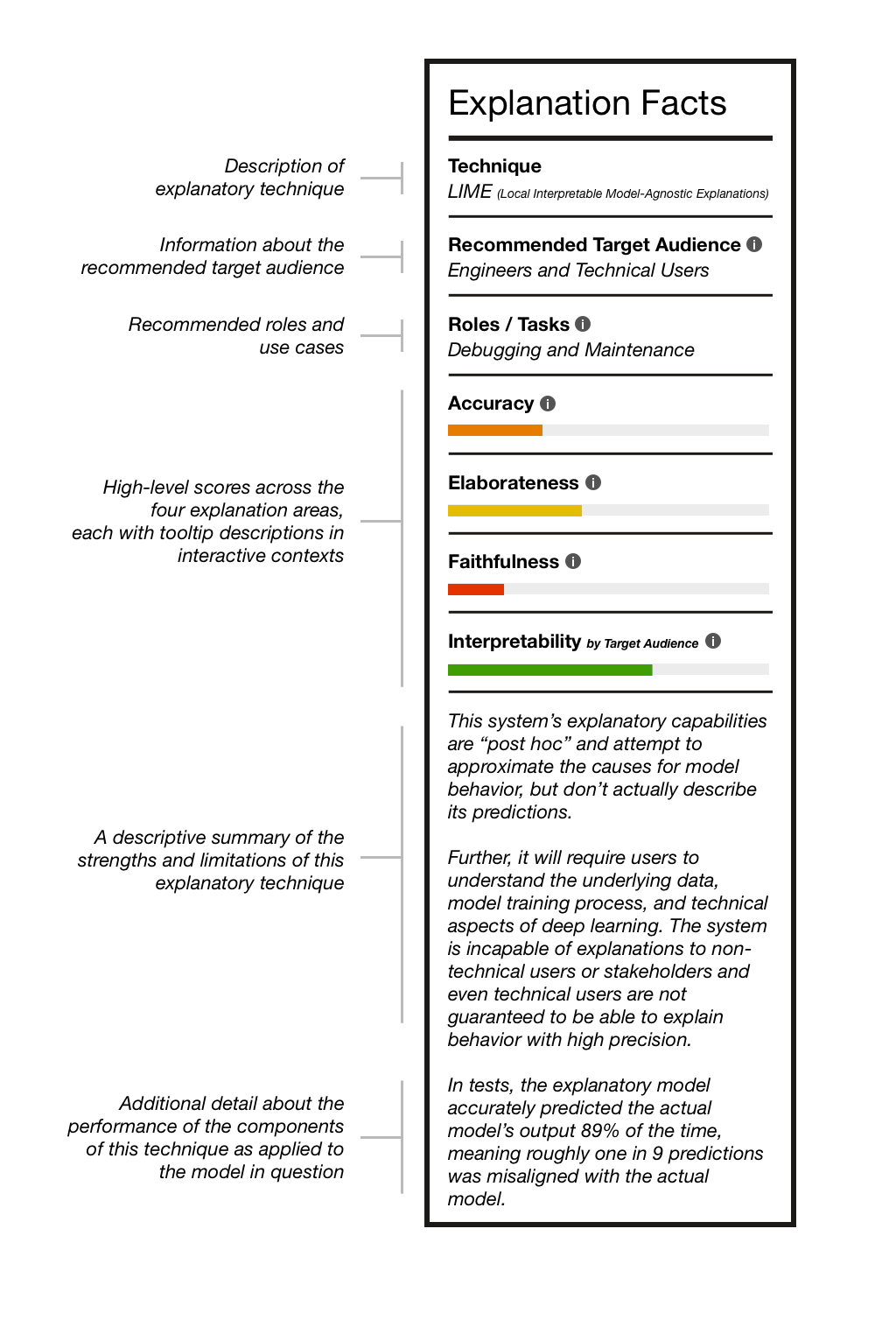}
    \caption{An example of a ``nutrition label'' for explanations with metrics for accuracy, elaborateness, faithfulness, and interpretability.  By using a diverse set of explanation criteria, the user or stakeholder is empowered to decide whether to trust the explanation (or not).}
    \label{fig:nutrition}
\end{figure}

\section{Discussion: Metrics - the next critical step for XAI}
The many colloquial meanings of ``explanation'' reflect the different functional roles they play. Because of the fluidity of what constitutes an explanation, we are faced with the possibility of building systems that provide explanations but not the sort that fit the need. They may seem ``acceptable'' but as has been demonstrated~\cite{langer1978mindlessness}, humans tend to accept even weak explanations if they have the right structural characteristics.

Explanations should provide new insights: explanations should go beyond the ``why, what, how''~\cite{foundations}.  We must also be able to compare and contrast them, as explanations can disagree and directly contradict each other~\cite{krishna2022disagreement}.   We extend these points and argue that developing metrics for explanations depends on both the audience as well as the capacity of the underlying explanation.  

Consider this metric of success: how well does the explanation actually ``explain'' or ``cover'' the space of decisions that the underlying (possibly opaque) system makes. This is our notion of faithfulness.  Imagine a set of explanations in the medical domain along with a faithfulness metric, e.g., ``the method that generated this explanation is 85\% faithful to the underlying model: 15\% of the time, the explanation will not explain the model faithfully.'' 

We also want to quantify the system capabilities: how well the explanation actually explains.  When we as humans explain something, it is a deductive process. Each claim follows from the last claim. We argue that the same process should be expected of machines and requires a delicate balance; explanations should be detailed, but not so much as to overload the target audience beyond their baseline knowledge. See Table~\ref{tab:targets} for an example.

The crucial point is that explanations need to fit the functional role in a manner that is interpretable to the audience. In evaluating an explanation, the question is whether it fits the role and the audience. Figure~\ref{fig:nutrition} a chart that maps functional roles onto the different aspects of an explanation that each requires. The question for any system is whether it can support generating explanations that fit each of these roles.  

With a variety of explanation types, a single ``one size fits all'' metric is inappropriate. Instead, we need a sort of nutrition label for explanations: a description of the intended purpose with a set of metrics that summarize the interpretability, accuracy, elaborateness, and faithfulness of any explanation with respect to the underlying goal and task. 

Explanations must be built, interpreted, and evaluated through the lens of the functions that they serve and the knowledge situations in which they do so; through the lens of requirements and a system’s capability to meet them. As work in XAI continues, it must attend to what exactly is being explained, to whom the explanation is directed, and how it is going to be used, and must provide a structured, standard evaluation approach via metrics of success.  
\begin{table}[]
    \centering
    \begin{tabular}{|c|c|c|c|c|}
    \hline
    \makecell{Target audience \\ and roles} & Interpretability & Accuracy & Elaborateness & Faithfulness  \\
    \hline 
    \hline
    \makecell{\textbf{Developers} \\ (Debugging)} & \makecell{Technical \\ Language: \\ Process/Data} & Process & \makecell{Decision Trees: \\ Process/Data} & \makecell{High \\(Debugging)} \\
    \hline
    \makecell{\textbf{Users} \\ (Interaction)}     & \makecell{Technical \\ Language: \\ Domain} & Domain & \makecell{Decision Features \\ Domain Features} & \makecell{Medium \\(Interaction)} \\
    \hline
    \makecell{\textbf{Stakeholders} \\ (Confidence)}     & \makecell{Stakeholder \\Goals} & Goals & Goals and risks & \makecell{Low \\(Confidence)} \\
    \hline
    \makecell{\textbf{Regulators} \\ (Responsibility)}     & \makecell{Design/Data \\ Decisions} & \makecell{System \\ Developer \\ Decisions} & \makecell{System/Developer \\ Decisions \\ and Audit trails} & \makecell{High \\(Responsibility)} \\
    \hline
    \end{tabular}
    \caption{A table documenting the target audience with the respective explanation capabilities that are ideal in terms of interpretability, accuracy, elaborateness, and faithfulness.}
    \label{tab:targets}
\end{table}

\section{Acknowledgements}
Funding for this work was provided by Underwriters Laboratories Inc. through the Center for Advancing Safety of Machine Intelligence.


\bibliography{main.bib}


\end{document}